\documentclass[11pt,showpacs]{revtex4}
\input epsf.sty

\usepackage{ulem}
\usepackage{cancel}
\usepackage{color}

\def\comment#1{}

\def\E{{\mathcal E}}

\usepackage{graphicx}
\begin{document}
\title{Resonant and nonresonant new phenomena of four-fermion operators for experimental searches}
\author{She-Sheng Xue}
\email{xue@icra.it}
\affiliation{ICRANeT, Piazzale della Repubblica, 10-65122, Pescara,\\
Physics Department, University of Rome ``La Sapienza'', Rome,
Italy} 
%\affiliation{$^{(b)}$Department of Physics, Isfahan University of Technology, Isfahan 84156-83111, Iran}

%\date{Received version \today}

\begin{abstract}
In the fermion content and gauge symmetry of the standard model (SM), 
we study the four-fermion 
operators in the torsion-free Einstein-Cartan theory.
The collider signatures of irrelevant operators are suppressed by the high-energy cutoff
(torsion-field mass) $\Lambda$, and cannot be experimentally accessible 
at TeV scales. Whereas the dynamics of relevant operators accounts for (i) 
the SM symmetry-breaking
in the domain of infrared-stable fixed point with the energy scale $v\approx 239.5$ GeV and (ii)
composite Dirac particles restoring the SM symmetry in the domain 
of ultraviolet-stable fixed point with the energy scale $\E\gtrsim 5$ TeV. 
To search for the resonant phenomena of composite Dirac particles 
with peculiar kinematic distributions in final states, we discuss possible high-energy 
processes: multi-jets and dilepton Drell-Yan process in LHC $p\,p$ collisions, 
the resonant cross-section in $e^-e^+$ collisions annihilating to hadrons 
and deep inelastic lepton-hadron $e^-\,p$ scatterings. To search for the  
nonresonant phenomena due to the form-factor of Higgs boson, 
we calculate the variation of Higgs-boson production and decay 
rate with the CM energy in LHC. We also present the discussions on 
four-fermion operators in the lepton sector and the mass-squared differences 
for neutrino oscillations in short baseline experiments.        
\end{abstract}

\pacs{12.60.-i,12.60.Rc,11.30.Qc,11.30.Rd,12.15.Ff}

\maketitle

%\vskip0.1cm
\noindent
{\bf Introduction.}
\hskip0.1cm
The parity-violating (chiral) gauge symmetries and spontaneous/explicit 
breaking of these symmetries for the hierarchy of fermion masses have 
been at the center of a conceptual elaboration that has played a major 
role in donating to mankind the beauty of the SM for 
particle physics. The Nambu-Jona-Lasinio (NJL) model \cite{njl} of 
dimension-6 four-fermion operators at high energies and its effective 
counterpart, the phenomenological model \cite{higgs} 
of elementary Higgs boson and its Yukawa-coupling to fermions 
at low energies, provide an elegant 
and simple description for the electroweak symmetry 
breaking and intermediate gauge boson masses. After a great 
experimental effort for many years, 
the ATLAS \cite{ATLAS} and CMS \cite{CMS} experiments have recently 
shown the first observations of a 126 GeV scalar particle in the search 
for the SM Higgs boson at the LHC. This far-reaching result
begins to shed light on this most elusive and fascinating arena of fundamental particle physics. 

In order to accommodate high-dimensional operators of fermion 
fields in the SM-framework of a well-defined quantum field theory at the 
high-energy scale $\Lambda$, it is essential and necessary to study: 
(i) what physics beyond the SM at the scale $\Lambda$ explains the 
origin of these operators; (ii) which dynamics of these operators 
undergo in terms of their couplings as functions of running energy 
scale $\mu$; (iii) associating to these dynamics where infrared (IR) 
or ultraviolet (UV) 
stable fixed point of physical couplings locates; (iv) in the domains 
(scaling regions) of these stable fixed points, which operators 
become physically relevant and renormalizable following renormalization 
group (RG) equations, 
and other irrelevant operators are suppressed by the cutoff at least 
$\mathcal O(\Lambda^{-2})$.    

The strong technicolor dynamics of extended gauge theories at the 
TeV scale was invoked \cite{hill1994,bhl1990a}
to have a natural scheme incorporating the 
relevant four-fermion operator 
$G(\bar\psi^{ia}_Lt_{Ra})(\bar t^b_{R}\psi_{Lib})$ of 
the $\langle\bar t t\rangle$-condensate model \cite{bhl1990}. 
On the other hand, these relevant operators can be constructed 
on the basis of phenomenology of the SM at low-energies.  
In 1989, several authors \cite{bhl1990,nambu1989,Marciano1989} 
suggested that the symmetry breakdown of the SM could be 
a dynamical mechanism of the NJL type that 
intimately involves the top quark at the high-energy scale $\Lambda$. 
Since then, many models based on this idea have been 
studied \cite{DSB_review}. The low-energy SM physics
was supposed to be achieved by the RG equations in the domain 
of the IR-stable fixed point 
with $v\approx 239.5$ GeV \cite{bhl1990a,Marciano1989,bhl1990}. 
In fact, the $\langle\bar t t\rangle$-condensate model 
was shown \cite{xue2013_1} to be energetically favorable, 
the top-quark and composite 
Higgs-boson masses are correctly 
obtained by solving RG equations in this IR-domain with the 
appropriate non-vanishing form-factor of Higgs 
boson in TeV scales \cite{xue2013,xue2014}.   

Inspired by the non-vanishing form-factor of Higgs boson, 
the formation of composite fermions
and restoration of the SM gauge symmetry in strong four-fermion 
coupling $G$ \cite{xue1997}, we preliminarily calculated 
the $\beta(G)$-function and showed  \cite{xue2014} the domain 
of an UV-stable fixed point at TeV scales, 
where the particle spectrum is completely different 
from the SM. This is reminiscent of the 
asymptotic safety \cite{w1} that quantum 
field theories regularized at UV cutoff $\Lambda$ 
might have a non-trivial 
UV-stable fixed point, RG flows 
are attracted into the UV-stable fixed point  
with a finite number of physically renormalizable operators. The weak and strong 
four-fermion coupling $G$ brings us into two distinct domains. 
This lets us recall the QCD dynamics: asymptotically free quark states in the domain of an UV-stable fixed point 
and bound hadron states in the domain of a possible IR-stable fixed point.

In this Letter, we proceed a further study on this issue, 
distinguishing physically relevant four-fermion operators from 
irrelevant one in the both domains of IR- and UV-stable fixed points, and 
focusing on the discussion of relevant operators and their 
resonant and nonresonant new phenomena for experimental searches. 

%\vskip0.1cm
\noindent
{\bf Four-fermion operators from quantum gravity.}
\hskip0.1cm
A well-defined quantum field theory for the SM Lagrangian requires a natural 
regularization (cutoff $\Lambda$) fully preserving the SM chiral-gauge 
symmetry. The quantum gravity provides a such regularization of discrete 
space-time with the minimal length $\tilde a\approx 1.2\,a_{\rm pl}$ \cite{xue2010}, 
where the Planck length 
$a_{\rm pl}\sim 10^{-33}\,$cm and scale $\Lambda_{\rm pl}=\pi/a_{\rm pl}\sim 10^{19}\,$GeV. 
However, the no-go theorem \cite{nn1981} tells us 
that there is no any consistent way to regularize 
the SM bilinear fermion Lagrangian to exactly preserve the SM chiral-gauge symmetry. 
This implies that the natural quantum-gravity regularization for the SM 
leads us to consider at least four-fermion operators. 

It is known that four-fermion operators of the classical and 
torsion-free Einstein-Cartan (EC) theory are naturally
obtained by integrating over ``static'' torsion fields at the Planck length,  
\begin{eqnarray}
{\mathcal L}_{EC}(e,\omega,\psi)&=&  {\mathcal L}_{EC}(e,\omega) + 
\bar\psi e^\mu {\mathcal D}_\mu\psi +GJ^dJ_d,
\label{ec0}
\end{eqnarray}
where the gravitational Lagrangian ${\mathcal L}_{EC}={\mathcal L}_{EC}(e,\omega)$, 
tetrad field $e_\mu (x)= e_\mu^{\,\,\,a}(x)\gamma_a$,
spin-connection field $\omega_\mu(x) = \omega^{ab}_\mu(x)\sigma_{ab}$,  
the covariant derivative ${\mathcal D}_\mu =\partial_\mu - ig\omega_\mu$ and 
the axial current $J^d=\bar\psi\gamma^d\gamma^5\psi$ of massless fermion fields. 
The four-fermion coupling $G$ relates to the gravitation-fermion gauge 
coupling $g$ and basic space-time cutoff $\tilde a$. 
In the regularized and quantized EC theory \cite{xue2010}
with a basic space-time cutoff, in addition to the leading term 
$J^dJ_d$ in Eq.~(\ref{ec0}) there are high-dimensional fermion operators ($d>6$), 
e.g., $\partial_\sigma J^\mu\partial^\sigma J{_\mu}$, which are suppressed
at least by ${\mathcal O}(\tilde a^4)$.

We consider massless left- and right-
handed Dirac fermions $\psi_{_L}$ and $\psi_{_R}$ carrying 
the SM quantum numbers, as well as
right-handed Dirac sterile neutrinos 
$\nu_{_R}$ and their Majorana counterparts 
$\nu^{\, c}_{_R}=i\gamma_2(\nu_{_R})^*$.  
Analogously to the EC theory (\ref{ec0}), we obtain a torsion-free, 
diffeomorphism and {\it local} gauge-invariant 
Lagrangian % (see for example Refs.~\cite{xue2010,art1989,contact}), 
\begin{eqnarray}
{\mathcal L}
&=&{\mathcal L}_{EC}(e,\omega)+\bar\psi_{_{L,R}} e^\mu {\mathcal D}_\mu\psi_{_{L,R}} 
+ \bar\nu^{ c}_{_{R}} e^\mu {\mathcal D}_\mu\nu^{ c}_{_{R}}\nonumber\\
&+&G\left(J^\mu_{_{L}}J_{_{L,\mu}} + J^\mu_{_{R}}J_{_{R,\mu}} 
+ 2 J^\mu_{_{L}}J_{_{R,\mu}}\right)\nonumber\\
&+&G\left(j^\mu_{_{L}}j_{_{L,\mu}} + 2J^\mu_{_L}j_{_{L,\mu}} 
+ 2 J^\mu_{_R}j_{_{L,\mu}}\right),
\label{art}
\end{eqnarray}
where we omit the gauge interactions in ${\mathcal D}_\mu$ 
and fermion flavor indexes of axial currents 
$J^\mu_{_{L,R}}\equiv \bar\psi_{_{L,R}}\gamma^\mu\gamma^5\psi_{_{L,R}}$ 
and $j^\mu_{_L}\equiv \bar\nu^c_{_R}\gamma^\mu\gamma^5\nu^c_{_R}$. 
The four-fermion coupling $G$ is unique for all four-fermion operators and 
high-dimensional fermion operators ($d>6$) are neglected. 
If torsion fields that couple to fermion fields 
are not exactly static, propagating a short distance 
$\tilde \ell \gtrsim \tilde a$, 
characterized by their large masses 
$\Lambda\propto \tilde \ell^{-1}$, this implies the four-fermion 
coupling $G\propto \Lambda^{-2}$.
\comment{
we will in future 
address the issue how the space-time cutoff $\tilde a$ due to quantum 
gravity relates to the scale $\Lambda(\tilde a)$ 
of four-fermion operators possibly by intermediate torsion fields 
or Kadonoff-Wilson renormalization group equation (RG).
In this Letter, 
we adopt the effective four-fermion operators (\ref{art}) 
in the context of a well-defined quantum field theory 
at the high-energy scale $\Lambda$.
}

In this article, we only discuss the relevance of 
dimension-6 four-fermion operators 
(\ref{art}), which can be written as 
\begin{eqnarray}
&+&(G/2)\left(J^\mu_{_{L}}J_{_{L,\mu}} + J^\mu_{_{R}}J_{_{R,\mu}} 
+ j^\mu_{_{L}}j_{_{L,\mu}} + 2J^\mu_{_L}j_{_{L,\mu}}\right)\label{art0}\\
&-&G\left(\, \bar\psi_{_L}\psi_{_R}\bar\psi_{_R} \psi_{_L}
+\, \bar\nu^c_{_R}\psi_{_R}\bar\psi_{_R} \nu^c_{_R}\right),
\label{art1}
\end{eqnarray}
by using the Fierz theorem \cite{itzykson}. Equations (\ref{art0}) and 
(\ref{art1}) represent repulsive and attractive operators respectively. 
It will be pointed out below that four-fermion operators (\ref{art0}) cannot be 
relevant and renormalizable operators of effective dimension-4 
in both domains of IR and UV-stable fixed points. We will consider only 
four-fermion operators (\ref{art1}) preserving the SM gauge symmetry 
without the flavor-mixing of three fermion families.

%\vskip0.1cm
\noindent
{\bf SM gauge symmetric four-fermion operators.}
\hskip0.1cm
In the quark sector, the four-fermion operators
\cite{xue2013_1} 
\begin{eqnarray}
G\left[(\bar\psi^{ia}_Lt_{Ra})(\bar t^b_{R}\psi_{Lib})
+ (\bar\psi^{ia}_Lb_{Ra})(\bar b^b_{R}\psi_{Lib})\right]+{\rm ``terms"},
\label{bhlx}
\end{eqnarray}
where $a,b$ and $i,j$ are the color and flavor indexes 
of the top and bottom quarks, the quark $SU_L(2)$ doublet 
$\psi^{ia}_L=(t^{a}_L,b^{a}_L)$ 
and singlet $\psi^{a}_R=t^{a}_R,b^{a}_R$ are the eigenstates 
of electroweak interaction. 
The first and second terms in Eq.~(\ref{bhlx}) are respectively 
the four-fermion operators of top-quark channel \cite{bhl1990} 
and bottom-quark channel,
whereas ``terms" stands for 
the first and second quark families that can be obtained 
by substituting $t\rightarrow u,c$ and $b\rightarrow d,s$. 

In the lepton sector, we 
introduce three right-handed sterile neutrinos $\nu^\ell_R$ 
($\ell=e,\mu,\tau$) that do not carry any SM quantum number. 
Analogously to Eq.~(\ref{bhlx}), the four-fermion operators 
in terms of gauge eigenstates are,
\begin{eqnarray}
G\left[(\bar\ell^{i}_L\ell_{R})(\bar \ell_{R}\ell_{Li})
+ (\bar\ell^{i}_L\nu^\ell_{R})(\bar \nu^\ell_{R}\ell_{Li}) 
+ (\bar\nu^{\ell\, c}_R\nu^{\ell\,}_{R})(\bar \nu^{\ell}_{R}\nu^{\ell c}_{R})\right],
\label{bhlxl}
\end{eqnarray}
preserving all SM gauge symmetries, 
where the lepton $SU_L(2)$ doublets $\ell^i_L=(\nu^\ell_L,\ell_L)$, singlets 
$\ell_{R}$ and the conjugate fields of sterile neutrinos 
$\nu_R^{\ell c}=i\gamma_2(\nu_R^{\ell})^*$.  
Coming from the second term in Eq.~(\ref{art1}), the last term in Eq.~(\ref{bhlxl}) 
preserves the symmetry 
$U_{\rm lepton}(1)$ for the lepton-number 
conservation, although $(\bar \nu^{\ell}_{R}\nu^{\ell c}_{R})$ 
violates the lepton number of family ``$\ell$'' by two units. 
Similarly, there are following four-fermion operators
\begin{eqnarray}
G\left[(\bar\nu^{\ell\, c}_R\ell_{R})(\bar \ell_{R}\nu^{\ell\, c}_{R})
+(\bar\nu^{\ell\, c}_R u^{\ell}_{a,R})(\bar u^{\ell}_{a,R}\nu^{\ell c}_{R})
+(\bar\nu^{\ell\,c}_R d^\ell_{a,R})(\bar d^{\ell}_{a,R}\nu^{\ell c}_{R})\right],
\label{bhlbv}
\end{eqnarray}
where quark fields $u^{\ell}_{a,R}=(u,c,t)_{a,R}$ and $d^{\ell}_{a,R}=(d,s,b)_{a,R}$.

In addition, there are SM gauge-symmetric four-fermion 
operators that contain quark-lepton interactions \cite{xue1999nu}, 
\begin{eqnarray}
G\left[(\bar\ell^{i}_Le_{R})(\bar d^a_{R}\psi_{Lia})
+(\bar\ell^{i}_L\nu^e_{R})(\bar u^a_{R}\psi_{Lia})\right] +~~{\rm ``terms"},
\label{bhlql}
\end{eqnarray}
where $\ell^i_L=(\nu^e_L,e_L)$ and $\psi_{Lia}=(u_{La},d_{La})$
for the first family. The ``terms'' represent   
for the second and third families with substitutions: 
$e\rightarrow \mu, \tau$, $\nu^e\rightarrow \nu^\mu, \nu^\tau$, and 
$u\rightarrow c, t$ and $d\rightarrow s, b$. 
Here we do not consider baryon-number violating operators. 
It would be interesting to study four-fermion operators in the 
framework of the $SU(5)$ or $SO(10)$ unification theory.

%\vskip0.1cm
\noindent
{\bf Relevant and irrelevant four-fermion operators in the IR-domain.}
\hskip0.1cm 
Based on the approach of large $N_c$-expansion with a fixed value 
$GN_c$, it is shown that in the domain (IR-domain) 
of IR-stable fixed point $G_cN_c$, where $N_c$ is the color number, the top-quark channel of 
operators (\ref{bhlx}) undergoes the NJL-dynamics of spontaneous symmetry 
breaking \cite{bhl1990}. As a result, the $\Lambda^2$-divergence 
(tadpole-diagram) is removed by the mass gap-equation, the 
top-quark channel of operators (\ref{bhlx}) becomes physically relevant and 
renormalizable operators of effective dimension-4. Namely, 
the effective SM Lagrangian with {\it bilinear} top-quark mass term and 
Yukawa-coupling to the composite Higgs boson $H$ at low-energy 
scale $\mu$ obey the RG equation approaching to
the low-energy SM physics characterized by the energy 
scale $v\approx 239.5$ GeV  \cite{bhl1990}
\begin{eqnarray}
L_{\rm eff} &=& L_{\rm kinetic} + L_{\rm gauge} + 
m_t \bar t t+ \bar g_{t}\bar t t H \nonumber\\ 
&+& \tilde Z_H|D_\mu H|^2-m_{_H}^2H^\dagger H
-\frac{\bar \lambda}{2}(H^\dagger H)^2.
\label{eff}
\end{eqnarray} 
The Dirac part of lepton-sector (\ref{bhlxl}) and (\ref{bhlbv}) does not undergo the 
NJL-dynamics of spontaneous symmetry breaking 
because its effective four-fermion coupling $(G_cN_c)/N_c$ is smaller 
than the critical value $(G_cN_c)$ \cite{xue2013_1}.  
Therefore, except the top-quark channel (\ref{eff}), 
all Dirac fermions are massless and 
four-fermion operators (\ref{art1}) are irrelevant 
dimension-6 operators, whose tree-level amplitudes of four-fermion scatterings are 
suppressed ${\mathcal O}(\Lambda^{-2})$ in the IR-domain. The hierarchy 
of fermion masses and Yukawa-couplings
is actually due to the {\it explicit symmetry breaking} 
and {\it flavor-mixing}, which were preliminarily studied \cite{xue1999nu,xue1997mx} by 
using the Schwinger-Dyson equation for Dirac fermion self-energy functions and
we will present some detailed discussions on this issue 
in the coming paper \cite{xue_2015}. 
 
On the other hand, in the domain (UV-domain)
of UV-stable fixed point $G_{\rm crit}>G_cN_c$, 
the phase transition takes place 
from the symmetry-breaking phase to the symmetric phase,  
the four-fermion operators (\ref{art1})
undergo the dynamics of forming composite 
fermions, e.g.~$[\bar\psi_{_L},(\bar\psi_{_R}\psi_{_L}) \psi_{_R}]$, 
preserving all SM gauge symmetries, and the characteristic energy 
scale is $\E\gtrsim 5$ TeV \cite{xue2014}. In the UV-domain, 
four-fermion operators (\ref{art1}) 
are expected to acquire anomalous dimensions and thus become relevant 
operators of effective dimension-4, in the sense of being 
renormalizable and obeying RG equations.   

However, in the both IR- and UV-domains, four-fermion 
operators (\ref{art0}) are irrelevant operators of dimension-6, and 
thus suppressed ${\mathcal O}(\Lambda^{-2})$ for the following reasons. 
In the IR-domain for low-energy SM physics, the four-fermion 
operators (\ref{art0}) are not associated with the NJL-dynamics 
of spontaneous symmetry breaking. In the UV-domain, four-fermion operators (\ref{art0}) 
are not associated with the dynamics of forming composite fermions, due to 
the absence of strong coupling limit that is a necessary condition to 
form three-fermion bound states \cite{xue1997}. 
As a consequence, the tree-level amplitudes 
of four-fermion scatterings represented by these 
irrelevant operators (\ref{art0}) are suppressed ${\mathcal O}(\Lambda^{-2})$. 

In fact, using dilepton events the ALTAS collaboration \cite{ATLAS2013} has carried out a 
searched for nonresonant new phenomena originating from contact interactions \cite{contact}  
(the left-left isoscalar model $J^\mu_{_{L}}J_{_{L,\mu}}$, which is commonly used as
a benchmark for this type of contact interaction searches \cite{pdg2012}), 
and shown no significant deviations 
from the SM expectation up to $\Lambda = \pi/\tilde a > 10$ TeV .

Moreover, it should be mentioned that in the IR domain, even taking into account 
the loop-level corrections to the tree-level amplitudes of four-fermion scatterings, the 
four-point vertex functions of irrelevant four-fermion operators in Eqs.~(\ref{art0}) 
and (\ref{art1}) are also suppressed by the cutoff scale $\Lambda$, thus their deviations 
from the SM are experimentally accessible.  
However, we recall that in both the IR- and UV-domains, four-fermion operators 
in Eqs.~(\ref{art0}) and (\ref{art1}) have loop-level contributions, 
via rainbow diagrams of two fermion loops, 
to the wave-function renormalization (two-point function) 
of fermion fields \cite{xue1997} and 
these loop-level contributions to the $\beta(G)$-function are negative \cite{xue2014}.

%\vskip0.1cm
\noindent
{\bf Nonresonant new phenomena of four-fermion operators.}
\hskip0.1cm 
In the IR-domain with the electroweak breaking scale 
$v=239.5$ GeV, 
the dynamical symmetry breaking of four-fermion 
operator $G(\bar\psi^{ia}_Lt_{Ra})(\bar t^b_{R}\psi_{Lib})$ of 
the top-quark channel (\ref{bhlx}) accounts for the masses of 
top quark, $W$ and $Z$ bosons as well as a Higgs boson composed 
by a top-quark pair ($\bar t t$) \cite{bhl1990}. 
It is shown \cite{xue2013,xue2014} that   
this mechanism consistently gives rise to the top-quark 
and Higgs masses of the SM (\ref{eff}) at low energies, 
provided the appropriate value of non-vanishing form-factor 
of composite Higgs boson at the high-energy scale 
$\E\gtrsim 5\,$ TeV.

\begin{figure}%[!h]
\begin{center}
\includegraphics[height=1.25in]{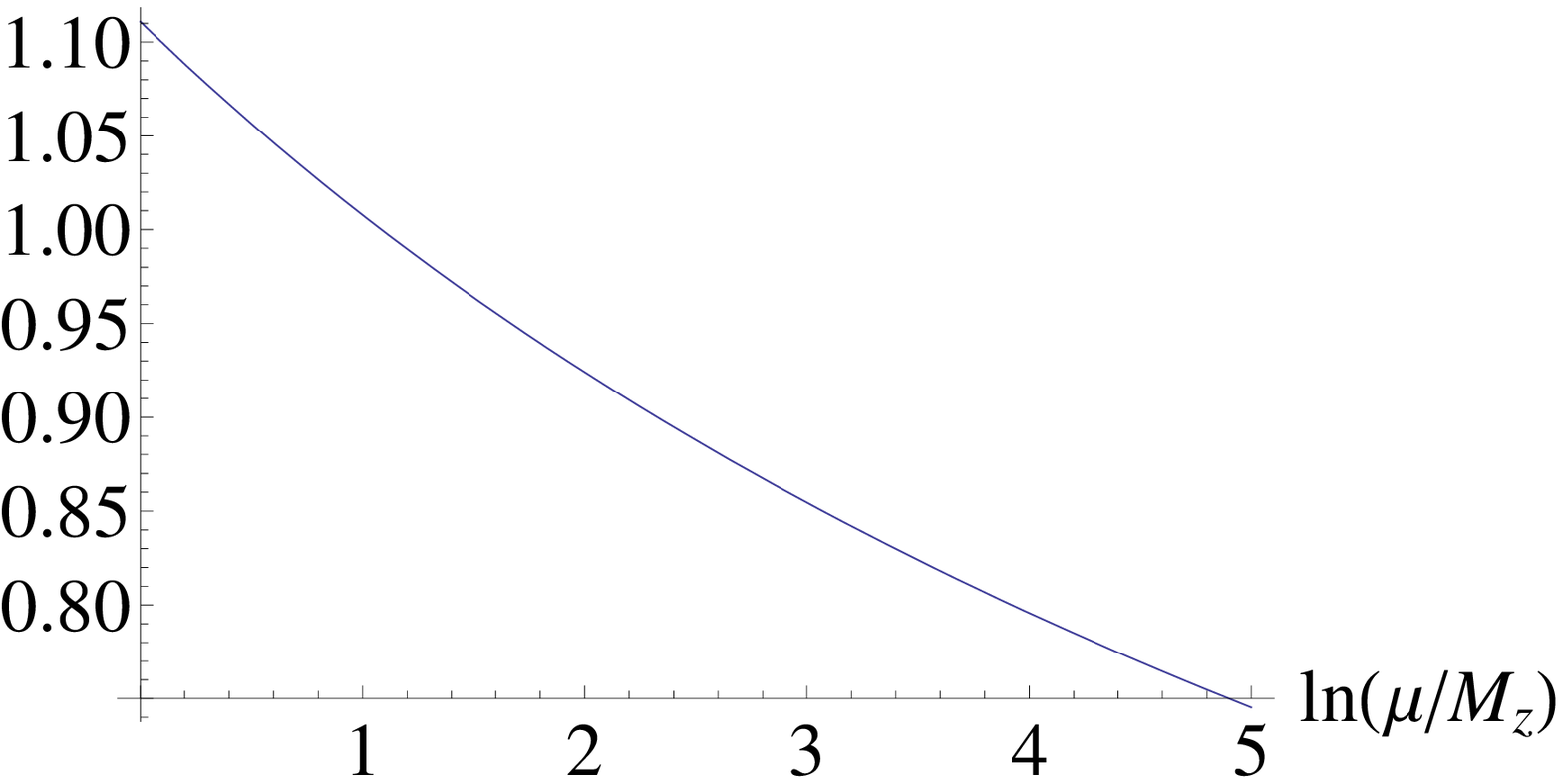}
\includegraphics[height=1.25in]{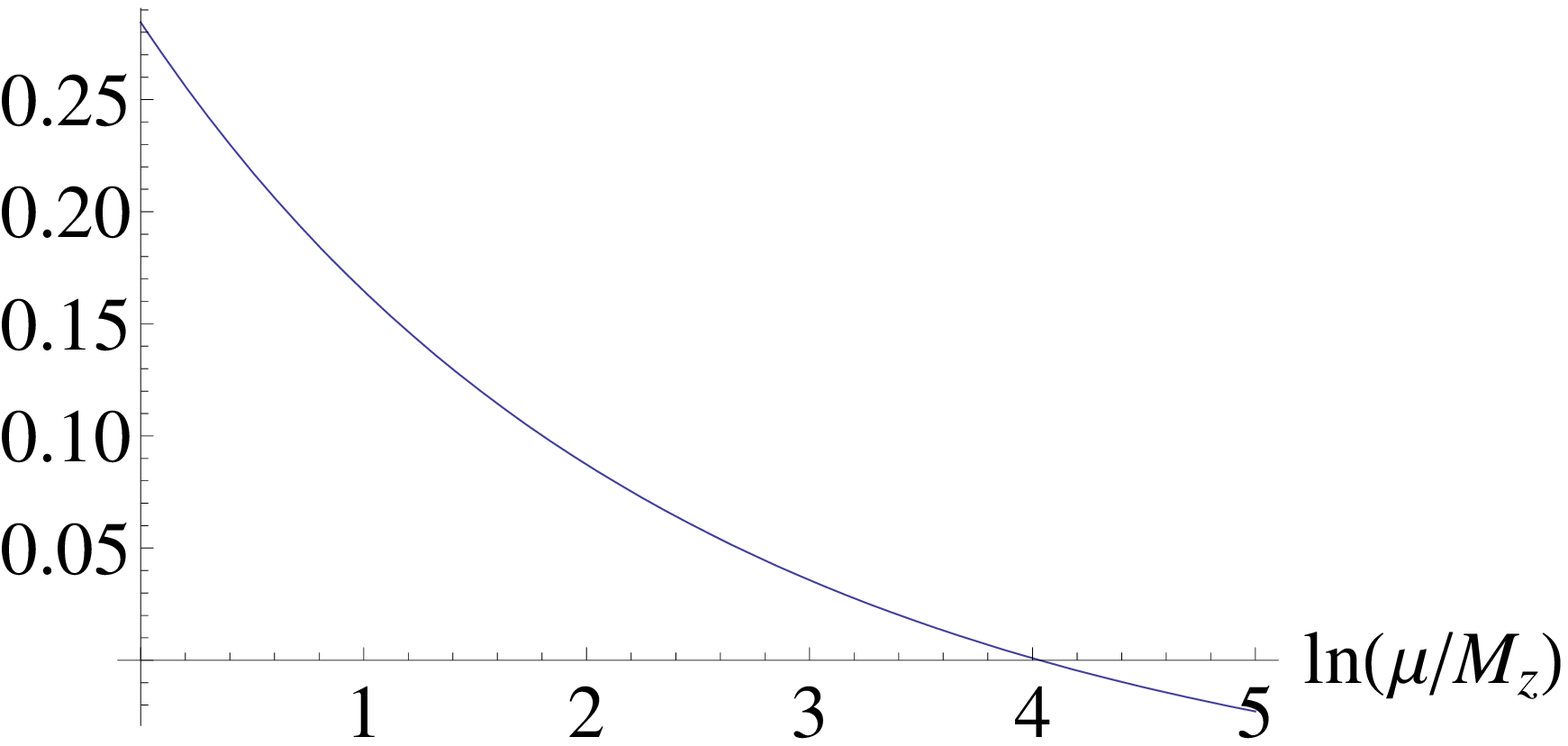}
\put(-370,95){\footnotesize $\bar g_t^2(\mu)$}
\put(-180,95){\footnotesize $\bar\lambda(\mu)$}
\caption{Using all experimentally measured SM quantities (including $m_t$ and $m_{_H}$) at low energies, we numerically solve the RG 
equations \cite{xue2014} to uniquely determine the  
Yukawa coupling $\bar g_t(\mu)$ of composite Higgs boson and top quark 
and the quartic coupling $\bar\lambda(\mu)$ of composite Higgs field 
in terms of the Higgs-boson energy scale $\mu> M_z$ up to 
$\E\approx 5$ TeV at which $\bar\lambda(\E)=0$.} \label{figyl}
\end{center}
\end{figure}

The finite form-factor of composite Higgs boson
is actually the wave-function renormalization of composite 
Higgs-boson field, which behaves as an elementary particle 
after performing the wave-function renormalization. 
However, the non-vanishing form-factor of composite Higgs boson 
is in fact related to the effective 
Yukawa-coupling of Higgs boson and top quark, i.e., 
$\tilde Z^{-1/2}_H(\mu)=\bar g_t(\mu)$ of Eq.~(\ref{eff}). 
The effective Yukawa coupling $\bar g_t(\mu)$
%\in [0.96, 0.89]$ for $\mu\in [m_t, \E]$
and quartic coupling $\bar\lambda(\mu)$ 
%\in [0.15, 0.0]$ for $\mu\in [m_{_H}, \E]$
monotonically decrease with the energy scale $\mu$ increasing 
in the range $m_{_H}< \mu <\E\approx 5$ TeV (see Fig.~\ref{figyl}). 
This means that the composite Higgs boson becomes more tightly bound as 
the  the energy scale $\mu$ increases.    

This should have some effects on the rates or cross-sections 
of the following three dominate processes of Higgs-boson production 
and decay \cite{ATLAS,CMS} or 
other relevant processes. Two-gluon fusion produces a Higgs boson via a top-quark loop, 
which is proportional to the effective Yukawa coupling $\bar g_t(\mu)$. Then,  
the produced Higgs boson decays to the two-photon state by coupling to a top-quark loop
and to the four-lepton state by coupling to two massive $W$-bosons or two massive $Z$-bosons.
Due to the $\bar t\,t$-composite nature of Higgs boson, the one-particle-irreducible (1PI) 
vertexes of Higgs-boson coupling to a top-quark loop, two massive $W$-bosons 
or two massive $Z$-bosons are proportional to the effective 
Yukawa coupling $\bar g_t(\mu)$. 
%The effective Yukawa coupling $\bar g^2_t(\mu)$
%\in [0.96, 0.89]$ for $\mu\in [m_t, \E]$ and quartic coupling $\bar\lambda(\mu)$ 
%\in [0.15, 0.0]$ for $\mu\in [m_{_H}, \E]$.  
As a result, both the Higgs-boson decaying rate to each of these three channels and 
total decay rate are proportional to $\bar g^2_t(\mu)$, 
which does not affect on the branching ratio of each Higgs-decay channel. 
The energy scale $\mu$ is actually the 
Higgs-boson energy, representing the total energy of final states, 
e.g., two-photon state and four-lepton states, to which the produced Higgs boson decays. 

These discussions imply that the resonant 
amplitude (number of events) of two-photon invariant 
mass $m_{\gamma\gamma}\approx 126$ GeV and/or 
four-lepton invariant mass $m_{4l}\approx 126$ GeV is expected 
to become smaller as the produced Higgs-boson energy $\mu$ increases, i.e., the
energy of final two-photon and/or four-lepton states increases,   
when the CM energy $\sqrt{s}$ of LHC $p\,p$ collisions increases 
with a given luminosity. 
Suppose that the total decay rate or each channel 
decay rate of the SM Higgs boson is measured at the Higgs-boson energy $\mu=m_t$ 
and the SM value of Yukawa coupling 
$\bar g^2_t(m_t)=2m_t^2/v\approx 1.04$ (see Fig.~\ref{figyl}). 
In this scenario of composite Higgs boson, as the Higgs-boson energy $\mu$
increases to $\mu=2m_t$, the Yukawa coupling 
$\bar g^2_t(2m_t)\approx 0.98$ (see Fig.~\ref{figyl}), 
the variation of total decay rate or each channel 
decay rate is expected to be 
$6\%$ for $\Delta \bar g^2_t\approx 0.06$. Analogously, the
variation is expected to be $9\%$ or $11\%$, at 
$\mu=3m_t$, $\bar g^2_t(3m_t)\approx 0.95$  or $\mu=4m_t$, 
$\bar g^2_t(4m_t)\approx 0.93$ (see Fig.~\ref{figyl}).
This variation may be still too small to be clearly 
distinguished by the present LHC experiments.  
Nevertheless, these effects are the nonresonant new signatures 
of low-energy collider that show the deviations of this scenario from the SM. 
It will be shown \cite{xue_2015} that the induced (1PI) Yukawa couplings 
$\bar g_b(\mu)$, $\bar g_\tau(\mu)$ and $\bar g_f(\mu)$ 
of composite Higgs boson to the bottom-quark, tau-lepton and other fermions 
also weakly decrease with increasing Higgs-boson energy, 
this implies a slight decrease of
number of dilepton events in the Drell-Yan process. 

However, the nonresonant new phenomena stemming 
from four-fermion scattering amplitudes of 
irrelevant operators of diemnsion-6 in 
Eqs.~(\ref{art0}) and (\ref{art1}) are suppressed 
${\mathcal O}(\Lambda^{-2})$ and hard to 
be identified in high-energy processes of 
LHC $p\,p$ collisions (e.g., the 
Drell-Yan dilepton process, see Ref.~\cite{ATLAS2013}), 
$e^-e^+$ annihilation to hadrons and deep inelastic lepton-hadron $e^-\,p$ 
scatterings at TeV scales. 

%\vskip0.1cm
\noindent
{\bf Composite particles in the UV-domain.}
\hskip0.1cm 
We turn to discuss the composite particle spectra and interacting 
vertexes in the UV-domain with the energy scale 
$ \E\gtrsim 5$ TeV. As the energy scale $\mu$ and four-fermion coupling 
$G(\mu)$ increase, composite Dirac fermions are formed  
at high energies $\mu > \E$ and strong coupling $G>G_{\rm crit}$.
For the top-quark channel, the composite Higgs boson 
combines with another top quark to form 
the massive composite Dirac fermions: $SU_L(2)
$ doublet ${\bf\Psi}^{ib}_D=(\psi^{ib}_L, {\bf\Psi}^{ib}_R)$ and singlet 
${\bf\Psi}^{b}_D=({\bf\Psi}^b_{L},t_R^b)$, where the renormalized composite 
three-fermion states are:
\begin{equation}
{\bf\Psi}^{ib}_R=(Z^{^S}_R)^{-1}(\bar\psi^{ia}_Lt_{Ra})t^b_{R}\,;
\quad {\bf\Psi}^b_{L}=(Z^{^S}_L)^{-1}(\bar\psi^{ia}_Lt_{Ra})\psi^{b}_{iL},
\label{bound}
\end{equation}
and the composite bosons ($SU_L(2)$-doublet) are 
$
H^i=[Z^{^S}_H]^{-1/2}(\bar\psi^{ia}_Lt_{Ra}),
$
where the form-factor $Z^{^S}_{R,L}$ and $[Z^{^S}_H]^{1/2}$ are 
generalized wave-function renormalization of composite fermion and boson operators. 
For the bottom-quark channel, the composite particles are represented by 
Eq.~(\ref{bound}) 
with the replacement $t_{Ra}\rightarrow b_{Ra}$, carrying the 
different quantum numbers of the $U_Y(1)$ gauge group. 
The same discussions also apply for the first and second quark families 
by substituting the $SU_L(2)$ doublet $(t_{La}, b_{La})$ 
into $(u_{La}, d_{La})$ or $(c_{La}, s_{La})$ and singlet 
$t_{Ra}$ into $u_{Ra}$ or $c_{Ra}$, as well as singlet 
$b_{Ra}$ into $d_{Ra}$ or $s_{Ra}$. Some detailed discussions can be found in 
Ref.~\cite{xue2014}.

Analogously, we present in this letter for the $e_R$-channel of quark-lepton 
interactions (\ref{bhlql}), 
the massive composite Dirac fermions: $SU_L(2)
$ doublet ${\bf\Psi}^{i}_D=(\ell^{i}_L, {\bf\Psi}^{i}_R)$ and singlet 
${\bf\Psi}_D=({\bf\Psi}_{L},e_R)$, where the renormalized composite 
three-fermion states are:
\begin{equation}
{\bf\Psi}^{i}_R=(Z^{^S}_R)^{-1}(\bar d^{\,a}_R \psi^i_{La})e_{R}\,;
\quad {\bf\Psi}_{L}=(Z^{^S}_L)^{-1}(\bar\psi^i_{La}d^{\,a}_R)\ell_{iL},
\label{boundem}
\end{equation}
and the composite bosons ($SU_L(2)$ doublet) are 
$
H^i=[Z^{^S}_H]^{-1/2}(\bar\psi^{i}_{La}d^{\,a}_{R}).
$
For the $\nu^{\,e}_R$-channel, the composite particles are represented 
by Eq.~(\ref{boundem}) with the replacements 
$e_{R}\rightarrow \nu^{\,e}_{R}$ and $d^{\,a}_{R}\rightarrow u^{\,a}_{R}$, 
\begin{equation}
{\bf\Psi}^{i}_R=(Z^{^S}_R)^{-1}(\bar u^{\,a}_R \psi^i_{La})\nu^{\,e}_{R}\,;
\quad {\bf\Psi}_{L}=(Z^{^S}_L)^{-1}(\bar\psi^i_{La}u^{\,a}_R)\ell_{iL},
\label{boundenu}
\end{equation}
and the composite bosons ($SU_L(2)$ doublet) are 
$
H^i=[Z^{^S}_H]^{-1/2}(\bar\psi^{i}_{La}u^{\,a}_{R}),
$
carrying
different quantum numbers of the $U_Y(1)$ gauge group. 
The composite particles from the second and third families 
can be obtained by substitutions: 
$e\rightarrow \mu, \tau$, $\nu^e\rightarrow \nu^\mu, \nu^\tau$, and 
$u\rightarrow c, t$ and $d\rightarrow s, b$. 

In addition, the composite three-fermion states formed by the first 
term ($\ell_R$-channel) of Eq.~(\ref{bhlxl}) are:
\begin{eqnarray}
{\bf\Psi}^{i}_R=Z^{^S}_R(\bar\ell^{i}_L\ell_{R})\ell_{R}\,;
\quad {\bf\Psi}_{L}=Z^{^S}_L(\bar\ell^{i}_L\ell_{R})\ell_{iL},
\label{boundl}
\end{eqnarray}
the composite Dirac fermions are $SU_L(2)$ doublet 
${\bf\Psi}^{i}_D=(\ell^{i}_L, {\bf\Psi}^{i}_R)$ and singlet 
${\bf\Psi}_D=({\bf\Psi}_{L},\ell_R)$, 
as well as the composite bosons ($SU_L(2)$ doublet) 
$H^i=[Z^{^S}_H]^{1/2}(\bar\ell^{i}_L\ell_{R})$. 
The composite particles formed by the second 
term ($\nu^\ell_R$-channel) of Eq.~(\ref{bhlxl}) are obtained by the 
replacement $\ell_R\rightarrow \nu^\ell_R$, and they carry no charge 
of the $U_Y(1)$ gauge group. These composite particles have 
the same mass and form factor as their counterparts of 
Eq.~(\ref{bound}) in the quark sector, but different 
quantum numbers of the SM gauge symmetries. 
 
In the UV-domain, all four-fermion operators (\ref{art1}) are expected to 
become relevant operators of effective dimension-4 by the formation of composite particles. 
The propagators of these composite particles have 
poles and residues that respectively 
represent their masses and form-factors.  
As long as their form-factors are finite, these composite particles 
behave as elementary particles with the mass-shell conditions
\begin{eqnarray}
E_{\rm com}=\sqrt{p^2+{\mathcal M}^2}\approx {\mathcal M},
\quad {\rm for }~~p\ll {\mathcal M}\approx \E_\xi
\label{mcp}
\end{eqnarray}
where the characteristic 
energy scale $\E_\xi$ sets in the UV-domain 
\begin{eqnarray}
\E_{\rm thre} \lesssim \E_\xi\ll \Lambda, \quad \E_{\rm thre}\gtrsim 5\, {\rm TeV}.
\label{scales}
\end{eqnarray} 
When the energy scale $\mu$
decreases to the energy threshold $\E_{\rm thre}$ and the four-fermion 
coupling $G(\mu)$ decreases to the critical value 
$G_{\rm crit}(\E_{\rm thre})$,  
all three-fermion and two-fermion bound states decay to 
their constituents of elementary particles of the SM. Needless to say, 
these composite particles should play important roles in the early 
universe and phase transition to the epoch of electroweak symmetry
breaking.

Compared with the SM in the IR-domain of the symmetry breaking phase, 
the spectrum of massive 
composite particles in the UV-domain of the symmetric phase 
has the same quantum numbers 
of SM chiral gauge symmetries, coupling to gauge bosons 
$\gamma,W^\pm, Z^0$ and gluon, but the spectrum and interacting vertex are 
vector-like, fully preserving the parity-symmetry. In the UV-domain 
of the symmetry phase, there is only the spectrum of composite fermions, 
as example the top quark channel (\ref{bound}), SM elementary 
fermions and Higgs boson become the constitutes of composite fermions.
It would be helpful to explain this scenario by analogy with the 
QCD of elementary quarks in high-energies and only composite hadrons 
in low energies. The $\beta_{\rm QCD}$-function has an opposite sign to 
the $\beta$-function of this scenario \cite{xue2014}.   
Similarly to that the composite 
Higgs behaves as an elementary particle in the IR-domain of the SM,
these composite fermions behave as elementary particles, provided
their wave-function renormalizations are finite above the threshold energy 
(\ref{scales}). Otherwise, composite particles dissolve (decay) to their constitutes 
of SM elementary particles.      
In Ref.~\cite{xue2014}, we discussed the resonant new phenomena (multi-jets events) 
of four-quark operators (\ref{bhlx}) and (\ref{bound}) with 
peculiar kinematic distributions in final states, that are expected to be observed 
in LHC $p\,p$ collisions.

%\vskip0.1cm
\noindent
{\bf Resonant new phenomena in high-energy processes.}
\hskip0.1cm 
In this section, we discuss that 
the resonant new phenomena of these composite particles
produced in high-energy processes: 
(i) the Drell-Yan process in LHC $p\,p$ collisions, 
(ii) $e^-e^+$ annihilation to hadrons 
and (iii) deep inelastic lepton-hadron $e^-\,p$ scatterings. 

(i) The Drell-Yan process: 
in addition to multi-jets produced in LHC $p\,p$ collisions \cite{xue2014}, the pair of 
lepton and anti-lepton are produced by the annihilation of quark and 
anti-quark via SM gauge bosons. The final states of 
lepton and anti-lepton (dilepton) are measured in terms of 
their invariant mass $m_{\ell\ell}$. In Ref.~\cite{ATLAS2013}, the 
analysis up to $m_{\ell\ell}\approx 1$ TeV has been made, showing no deviation from the SM 
attributed to the four-fermion operators (contact interactions) $J^\mu_{_{L}}J_{_{L,\mu}}$ 
in Eq.~(\ref{art0}), which are in fact irrelevant and suppressed $O(\Lambda^{-2})$ as 
previously discussed. 

Instead, the relevant four-fermion operators (\ref{bhlql}) 
form massive composite fermions of Eqs.~(\ref{boundem}) and (\ref{mcp}), i.e., 
the $e$-channel
$[e_L, (\bar d^{\,a}_R d_{La})e_{R}]$ 
and/or the $\mu$-channel  $[\mu_L,(\bar s^{\,a}_R s_{La})\mu_{R}]$, 
which then decay to lepton and anti-lepton as final states. 
This implies the resonant new phenomena to appear in the dilepton 
invariant mass $m_{\ell\ell}\approx {\mathcal M}\gtrsim 5$ TeV. 
Due to their origin from a very massive 
composite particle ``at rest'', the kinematic distribution of dilepton final states 
is expected to be two leptons moving apart in opposite directions, each carrying 
energy-momentum about ${\mathcal M}/4$. 
In the channels involving neutrino or sterile neutrinos, 
the four-fermion operators (\ref{bhlql}) form the composite fermions 
of Eqs.~(\ref{boundem},\ref{boundenu}) and (\ref{boundl}), which then decay 
to final states containing 
neutrinos $\nu_L$ and/or sterile neutrinos $\nu_R$. These neutrinos 
carry away some missing transverse energy momenta, 
which is an important signal (or trigger) 
for a number of new phenomena. 

(ii) Deep inelastic lepton-hadron scatterings (DIS):
the deep inelastic electron $e(k)$ and hadron $N(p)$ scattering
$e(k)+N(p)\rightarrow e(k')+ X(p_n)$ occurs via exchange of 
neutral gauge bosons in the SM, where the energy transfer 
$q=k-k'$ and $X(p_n)$ represents some hadronic final states 
with total four-momentum $p_n$. 
If the energy transfer $q>{\mathcal M}\gtrsim 5$ TeV, the composite Dirac
fermion $[e_L, (\bar d^{\,a}_R d_{La})e_{R}]$ formed by the relevant 
four-fermion operator (\ref{bhlql}). This implies the appearance of a 
resonant cross-section $\sigma_{e\,p}(q)$ whose peak locates at $q\approx{\mathcal M}$ 
the mass of the composite Dirac fermion formed intermediately. In addition, the kinematic 
distribution of final states $e(k')$ and $X(p_n)$ is expected to be deviated 
from the SM result. In the channels involving neutrinos $\nu_L$ 
and $\nu_R$ of Eq. (\ref{bhlql}), the composite Dirac
fermions $[e_L, (\bar u^{\,a}_R d_{La})\nu^e_{R}]$ 
and $[\nu^e_L, (\bar d^{\,a}_R u_{La})e_{R}]$ form and decay in the 
electron $e(k)$ and hadron $N(p)$ scattering process, the kinematic 
distribution of final states $e(k')$ and $X(p_n)$ is expected to be affected by  
the missing energy-momenta carried away by neutrinos.

In addition, the polarized electron-deuteron deep inelastic (DIS) 
experiment \cite{pdg2012,slacdis} measured the right-left asymmetry,
\begin{eqnarray}
A=\frac{\sigma_{_R}-\sigma_{_L}}{\sigma_{_R}+\sigma_{_L}}
\label{dis}
\end{eqnarray} 
where $\sigma_{_{R,L}}(q)$ is the cross-section for the deep-inelastic 
scattering of a right- or left-handed electron 
$e_{_{R,L}}+N\rightarrow e +X$. For low energy-momentum transfer 
$q^2\ll \E$, $A\not=0$ for the violation of parity symmetry of the SM. 
As high energy-momentum transfer $q$ approaches $\E$
and $q >\E$, $A\rightarrow 0$ for the restoration of parity symmetry \cite{xue2013}.

(iii) $e^+e^-$ collisions annihilating to hadrons:
the process and cross-section of electron-positron annihilation to hadrons
have been important to study the SM.
As the CM energy $\sqrt{s}$ (invariant mass) 
of electron-positron collider increases, 
the bound states of quarkonia/mesons have been 
discovered as narrow resonances in the cross-section, 
the resonance of $Z$-production has been also observed for the CM energy
$\sqrt{s}\gtrsim 100\,$GeV \cite{pdg2012}. The narrow resonance for 
$(\bar t t)$ quarkium  will appear in final states if $\sqrt{s}> 346\,$GeV. On the basis of relevant four-fermion operators (\ref{bhlxl}) and 
(\ref{bhlql}), we expect the resonances of composite Dirac 
particles, e.g.,  $[\bar e^{}_{L},(\bar e^{}_Le^{}_{R})e^{}_{R}]$, 
$[\bar \nu^{\,e}_{L},(\bar e^{}_{R}\nu^{\,e}_L)e^{}_{R}]$, 
$[\bar e^{}_{L},(\bar \nu^{\,e}_Re^{}_{L})\nu^{\,e}_{R}]$, 
$[\bar e^{}_{L},(\bar d^{\,a}_Rd^{}_{La})e^{}_{R}],\cdot\cdot\cdot$, 
to appear in the cross-section of $e^+\,e^-$ annihilation, 
if the CM energy of the electron-positron collider reaches the energy 
threshold $\E_{\rm thre}$, i.e., 
$\sqrt{s}\gtrsim\E_{\rm thre}\gtrsim 5$ TeV. However, this energy scale 
does not seems to be reached by both $e^+e^-$ colliders and DIS experiments in near future.
%The cross-section of producing a composite Dirac particle, which then decays, should be proportional to its inverse squared mass $\sim 1/{\mathcal M}^2$.

In addition, two high-energy photons from the LHC $pp$ 
collision can produce two electron-positron pairs fusing 
into a composite Dirac particle (\ref{boundl},\ref{mcp}) 
with $\ell^i_L=e_L$ and $\ell_R=e_R$, which can be 
identified by observing the resonance in the invariant 
mass ${\mathcal M}_{e^+e^-}$ of final states of two electron-positron pairs. 
In the CM frame, electron and positron of each pair 
move apart in the opposite direction and the energy-momentum of 
each particle is about one-fourth of the invariant mass. 
The cross-section for these channels can be estimated to be 
$4\pi \alpha^2 /{\mathcal M}^2$.
The massive composite particle (\ref{boundl},\ref{mcp}) that comprises electron 
$\ell^i_L=e_L$ and sterile neutrino $\ell_R=\nu_R$ 
is expected to be identified by observing the resonance with final states of 
electron and positron oppositely moving apart with energy being 
one-half of the invariant mass ${\mathcal M}$, and the rest carried away by sterile 
neutrino and anti-neutrino oppositely moving apart. Compared with 
the multi-jets resonant phenomena due to four-quark operators (\ref{bound})  
in LHC $p\,p$ collisions \cite{xue2014}, the probabilities of 
the dilepton resonant phenomena due to four-lepton operators (\ref{boundl}) 
in $p\,p$ collisions are smaller, because of the factor $\alpha^2$.

In general, what can be said are following.  
If the accessible CM energy $\sqrt {s}>{\mathcal M}$, 
the cross section for the allowed
inelastic processes forming massive 
composite Dirac particles will be geometrical
in magnitude, of order $\sigma_{\rm com}\sim 4\pi /{\mathcal M}^2$ in 
the CM frame where massive composite Dirac particles are approximately at 
rest. Decays of these massive 
composite particles to their constituents leading to unconventional 
events of multi-jets, jets-dilepton and multi-leptons states
with peculiar kinematic distributions in the CM frame. 
As a result, these unconventional events will qualitatively depart from 
the SM, and completely dominate over the SM processes, 
for which cross sections go roughly as $\pi \alpha^2_{\rm gauge}/(\sqrt{s})^2$. Thus the 
SM background is expected to be more or less zero.
On the other hand, if the accessible CM energy $\sqrt {s}<{\mathcal M}$,
then departures from the SM will be quantitative rather 
than qualitative, as described  
in the previous section.
 
In currently scheduled LHC ($p\,p$-collision) runs for next 20 years, 
the integrated luminosity will go from 
$10\,{\rm fb}^{-1}$ up to $10^{3}\,{\rm fb}^{-1}$ and the CM energy 
$\sqrt{s}\,$ from $7
$ TeV up to $14$ TeV, then the number of events of
composite Dirac particles created in quark-quark or 
quark-lepton (Drell-Yan) processes can be estimated by 
$\sigma_{\rm com}\times  10^{1-3}{\rm fb}^{-1}\sim 4\pi\times 10^{5-7}$ 
for the $(u,d)$ family, $\sim 4\pi\times 10^{3-5}$ for the $(c,s)$ 
and $(t,b)$ families, assuming ${\mathcal M}\sim 5\,$TeV. 
%The background of signals attributed to the SM physics has been studied (see for example \cite{zhang_paper}). 

%\vskip0.1cm
\noindent
{\bf Neutrino sector.}
\hskip0.1cm 
In the UV-domain, 
the four-neutrino operator ($\nu^{\ell c}_R$-channel) of the last term of 
Eq.~(\ref{bhlxl}) forms composite three-fermion states 
(self-conjugate Majorana states)
\begin{eqnarray}
{\bf\Psi}^{\ell}_R=Z^{^S}_R(\bar\nu^{\ell\,' c}_R\nu^{\ell\, '}_{R})\nu^{\ell}_{R}\,;
\quad {\bf\Psi}^{\ell c}_R=Z^{^S}_L(\bar\nu^{\ell\,'c}_R\nu^{\ell\,'}_{R})\nu^{\ell c}_R,
\label{boundm}
\end{eqnarray}
and composite Majorana fermions
${\bf\Psi}^{\ell}_M=(\nu^{\ell c}_R, {\bf\Psi}^{\ell}_R)$, 
as well as the composite bosons 
$H_M=[Z^{^S}_H]^{1/2}(\bar\nu^{\ell\,'c}_R\nu^{\ell\,'}_{R})$. 
These composite particles comprising only neutrinos are hard to be produced 
in ground laboratories, except in the early universe.

However, in the IR-domain for the SM, 
the last term of Eq.~(\ref{bhlxl}) can 
generate a mass term of Majorana type, because the family number 
$N_i=3$ ($i=1,2,3$ that are different from the $SU_L(2)$ index) 
plays the same role as the color number $N_c$ of 
the top-quark in the $\langle \bar tt\rangle$-condensate. 
In the same way similar to the gauge and mass eigenstates of neutrinos 
$\nu^{\ell}_L=U_{\ell\, i}\nu^i_L$, we use the $3\times 3$
unitary PMNS matrix $U$ for neutrino mixings to define the flavor eigenstates of 
sterile neutrinos by 
\begin{equation}
\nu^{\ell}_R=U_{\ell\, i}\nu^i_R,\quad ( 
\ell=e,\mu,\tau,\, i=1,2,3).
\label{rtan}
\end{equation}
Analogously to the generation of top-quark mass $m_t$, 
the dynamical symmetry breaking of the 
$U_{\rm lepton}(1)$-symmetry generates the Majorana mass of 
right-handed neutrinos 
\begin{eqnarray}
m^M=-G\sum_i\langle\bar\nu^{i\, c}_R\nu^{i\,}_{R}\rangle
\label{mam}
\end{eqnarray}
together with a sterile massless Goldstone boson, i.e.~the bound 
state ($\bar\nu^{i\,c}_R\gamma_5\nu^i_R$), and 
a sterile massive scalar particle, i.e.~the bound state 
$(\bar\nu^{i\,c}_R\nu^i_R)$, carrying two units of the lepton number 
of the family ``$i$''. Note that the family index ``$i$'' 
is summed over as the color index ``$a$'' in the 
$\langle\bar tt\rangle$-condensate.  
In the IR-domain, 
the sterile neutrino mass $m^M$ and sterile scalar 
particle mass $m^M_{_H}$ satisfy the mass-shell conditions, 
\begin{eqnarray}
m^M=\bar g_t(m^M)v_{\rm sterile}/\sqrt{2},\quad (m^M_{_H})^2/2
=\tilde \lambda (m^M_{_H}) v_{\rm sterile}^2,
\label{smass}
\end{eqnarray}
where $\bar g_t(\mu^2)$ and $\tilde \lambda (\mu^2)$ obey the same 
RG equations (absence of gauge interactions) 
and boundary conditions of Eqs.~(7,8,9) and (10) 
in Ref.~\cite{xue2014}.  
However, unlike the electroweak scale $v$ determined by the 
gauge-boson masses $M_{_W}$ and $M_{_Z}$ experimentally 
measured, the scale $v_{\rm sterile}$ 
is unknown and needs to be determined by the sterile neutrino 
mass $m^M$. The ratio is approximately estimated
\begin{equation}
\frac{m^M}{m^M_{_H}}=\frac{\bar g_t(m^M)}{2[\tilde\lambda(m^M_{_H})]^{1/2}}
\approx \frac{m_t}{m_{_H}}=\frac{\bar g_t(m_t)}{2[\tilde\lambda(m_{_H})]^{1/2}} = 1.37.
\label{rationu}  
\end{equation}
All sterile particles and gauge-singlet (neutral) states of  
massive composite Dirac particles (TeV-scales) discussed here 
can be possible candidates of warm and cold 
dark matter \cite{xuejpg2003}, and they can decay into SM particles via relevant 
four-fermion operators in Eqs.~(\ref{bhlx}-\ref{bhlql}).

As a result, in the IR-domain, 
we write the following bilinear Dirac and Majorana mass terms of neutrinos 
in terms of their mass eigenstates $\nu^i_L$ and $\nu^i_R$ 
\begin{eqnarray}
m^{D}_i\bar\nu^{i}_L\nu^i_{R}+ m^{M}_i\bar \nu^{i c}_{R}\nu^{i}_{R}+ {\rm h.c.}
\label{lmass}
\end{eqnarray}
We expect the Majorana masses to be approximately equal (degenerate), 
$m^{M}_i\approx m^{M}$, for three right-handed neutrino $\nu_R^i$, 
since there is not any preferential 
$i$-th component of the condensate 
$\langle\bar\nu^{i\, c}_R \nu^{i}_{R}\rangle$. 
Whereas, due to the origin from the explicit symmetry breaking 
terms related to family flavor mixings, the Dirac masses $m_i^D$ 
have the structure of 
hierarchy \cite{xue1999nu,xue1997mx,xue_2015}. Moreover 
we assume that $ m^M_i\gg m_i^D$.

Following the usual approach \cite{zli},  diagonalizing 
the $2\times 2$ mixing matrix (\ref{lmass}) in terms
of the neutrino and sterile neutrino mass eigenstate ``$i$'', 
we obtains the mixing angle $2\theta_i =\tan ^{-1} (m^D_i/m^M_i)$   
and two mass eigenvalues
\begin{eqnarray}
M_i^{\pm}= \frac{1}{2} \Big\{m^M_i\pm \left[(m^M_i)^2 
+ (m^D_i)^2\right]^{1/2}\Big\},\quad i=1,2,3,
\label{2lmass}
\end{eqnarray}
corresponding to two mass eigenstates: three heavy sterile Majorana 
neutrinos ($\nu^i_R+\nu_R^{ic}$) of approximately degenerate mass-spectra 
$M_i^{+}\approx m^M_i\approx m^M$, and three light gauge Majorana neutrinos 
($\nu^i_L+\nu_L^{ic}$) of hierarchic mass-spectra 
$M_i^{-}\approx (m^D_i)^2/4m^M_i$. The mixing angle 
$2\theta_i \approx (m^D_i/m^M_i)\ll 1$ and mass-squared difference 
$(M_i^{+})^2-(M_i^{-})^2\approx (m^M_i)^2$, indicating that the mixing of 
gauge and sterile Majorana neutrinos is very small. 
Therefore the oscillation between gauge and sterile Majorana neutrinos
can be negligible. 

We turn to discuss the neutrino flavor oscillations in the usual framework. 
The mass-squared difference of neutrino mass eigenstates ($i, j=1,2,3$),  
\begin{eqnarray}
\Delta M^2_{ij}\equiv (M_i^{\pm})^2-(M_j^{\pm})^2 \approx  \frac{1}{4} 
(2\Delta m^{2M}_{ij} +  \Delta m^{2D}_{ij} ),
\label{2mass}
\end{eqnarray}
where $\Delta m^{2M,D}_{ij}\equiv (m^{M,D}_{i})^2-(m^{M,D}_{j})^2$ 
[see Eq.~(\ref{lmass})]. 
Since $\Delta m^{2D}_{ij}\gg \Delta m^{2M}_{ij}$, the neutrino mass-squared difference
$\Delta m^{2M}_{ij}$ accounts for  
neutrino flavor oscillations with $E_\nu/L\sim\Delta m^{2M}_{12}\simeq 5
\times 10^{-3}{\rm eV^2}$ in the long-baseline experiments, where $E_\nu$ and $L$ 
respectively are neutrino energy and travel distance from a 
source to a detector.
Whereas the neutrino mass-squared difference 
$\Delta m^{2D}_{ij}$ may accounts for neutrino 
oscillations $E_\nu/L\sim\Delta m^{2D}_{ij}\gg 10^{-3}{\rm eV^2}$ in short 
baseline experiments \cite{Aguilar:2001ty}. 

%\vskip0.1cm
\noindent
{\bf Some remarks.}
\hskip0.1cm 
The multitude of seemingly arbitrary
parameters required to specify the SM shows the incompleteness of the SM, which 
is mainly manifested by our ignorance of (i) the relevant operators 
and dynamics that underlie the spontaneous/explicit breaking 
of the SM chiral-gauge symmetries, (ii) the global symmetries and mixings 
of puzzling replication of quark and lepton families. The relevant
four-fermion operators (\ref{art1}) potentially give the 
theoretical description of the SM in the IR-domain with $v\approx 239.5$ GeV, 
provided the UV-domain with $\E\gtrsim 5$ GeV, where the 
resonant and nonresonant new phenomena are distinct from the SM.     
We advocate that it is deserved to theoretically study the particle 
spectrum and symmetry of the strong-coupling theory (\ref{art1}) in the 
domain of UV-stable fixed point by non-perturbative numerical 
simulations, meanwhile experimentally search for resonant and 
nonresonant new phenomena of relevant four-fermion operators (\ref{art1}). 

%\vskip0.1cm
\noindent
{\bf Acknowledgment.}  
\hskip0.1cm Author is grateful to 
Prof.~Zhiqing Zhang for discussions on the LHC physics, and
Prof.~Hagen Kleinert for discussions on 
the IR- and UV-stable fixed points of quantum field theories.

\end{document}